\documentclass[12pt,preprint]{aastex} 

\shorttitle{GLOBAL TWIST OF SUNSPOT MAGNETIC FIELDS}
\shortauthors{S. K. Tiwari et al.}

\begin{document}

\title{GLOBAL TWIST OF SUNSPOT MAGNETIC FIELDS OBTAINED FROM
HIGH RESOLUTION VECTOR MAGNETOGRAMS}

\author{SANJIV KUMAR TIWARI and
        P.~VENKATAKRISHNAN}
\affil{Udaipur Solar Observatory, Physical Research Laboratory,
 Dewali, Bari Road, Udaipur-313 001, India}
\email{stiwari@prl.res.in}
\email{pvk@prl.res.in}
\and
\author{K. SANKARASUBRAMANIAN}
\affil{Space Astronomy \&\ Instrumentation Division, ISRO Satellite
Center, Airport Road, Vimanapura, Bangalore-560017, India}
\email{sankar@isac.gov.in}
\begin{abstract}
The presence of fine structures in the sunspot vector
magnetic fields has been confirmed from Hinode as well
as other earlier observations.
We studied 43 sunspots based on the data sets taken
from ASP/DLSP, Hinode (SOT/SP) and SVM (USO).
In this \emph{Letter}, (i) We introduce the concept of signed
shear angle (SSA) for sunspots and establish its
importance for non force-free fields.
(ii) We find that the sign of global $\alpha$
(force-free parameter)
is well correlated with the global SSA and the
photospheric chirality of sunspots.
(iii) Local $\alpha$ patches of opposite signs are present
in the umbra of each sunspot. The amplitude of the
spatial variation of local $\alpha$ in the umbra
is typically of the order of the global
$\alpha$ of the sunspot.
(iv) We find that the local $\alpha$ is distributed
as alternately positive and negative filaments in the
penumbra. The amplitude of azimuthal
variation of the local $\alpha$ in the penumbra is
approximately an order of magnitude larger than
that in the umbra. The contributions of the local
positive and negative currents and $\alpha$ in the
penumbra cancel each other giving almost no
contribution for their global values for whole sunspot.
(v) Arc-like structures (partial rings) with a
sign opposite to that of the
dominant sign of $\alpha$ of the umbral region
are seen at the umbral-penumbral
boundaries of some sunspots.
(vi) Most of the sunspots studied, belong to
the minimum epoch of the 23$^{rd}$ solar cycle and do
not follow the so-called hemispheric helicity rule.

\end{abstract}


\section{INTRODUCTION}
Helical patterns in the sunspots and associated features have
been observed for a long time \citep{hale25,hale27,rich41}
with a hemispheric preference of their chirality, which is
independent of solar cycle. Since the 90's, the subject has
been intensively revisited and the similar behaviour of
hemispheric patterns for various solar features have been
reported by many researchers
(\cite{hagi04,Nand06,bern05,pevt01,pevt07} and references
therein). However, this hemispheric behaviour
needs further investigation due to some inconsistencies
reported for different phases of a solar cycle and also for
data sets obtained from different
magnetographs \citep{hagi05,pevt08}.

For a force-free field, the global twist per unit axial length
is given by the force-free parameter $\alpha$ (see Appendix A
of \cite{tiw09}).
Some recent studies \citep{tiw08,tiw09a,tiw09b} have shown
that the global $\alpha$ of an active region bears the same sign as
its associated features/structures observed at chromospheric and
coronal heights. The chromospheric and coronal sign of twist is
inferred from the topological chirality sign of the observed
features. This leads us to believe that some form of
the photospheric global twist exists on the scale of sunspots.
However, the structures in the sunspot fields revealed by
modern vector magnetographs with high spatial and spectral
resolution compels us to make a careful revaluation of
global $\alpha$ and its physical meaning.

Since the photospheric field is not force-free \citep{metc95},
we need an alternative measure of the twist other than $\alpha$.
We introduce the concept of signed shear angle (SSA) for sunspot
magnetic fields in this paper and show how SSA is directly
related to vertical current ($J_z$) and $\alpha$, irrespective
of the force-free nature of the sunspot fields.

The presence of oppositely directed currents in
a single unipolar sunspot was first shown by \cite{sever65}.
For a detailed investigation of local $\alpha$ distribution
in three sunspots using 46 vector magnetograms see
\cite{pcm94}.
Recently, \cite{su09} reported an interesting pattern of
fine structures in the $\alpha$ distribution within one
active region (AR) using Hinode data with higher resolution.
We present a comprehensive study of 43 sunspots with high
resolution and establish the contribution
of such fine structures to the global twist. For this purpose
we will rely on $J_z$ and $\alpha$ values.

The helicity hemispheric rule or more precisely twist
hemispheric rule is claimed to be established
by many researchers \citep{seeh90,pcm95,abra96,bao98,long98,hagi05}
and has recently been a matter of some debate
\citep{hagi05,pevt08}.
A model developed by \cite{chou04} predicts deviation
from the twist hemispheric rule in the beginning of the
solar cycle. However, some observers claim this deviation
from the hemispheric rule may be present in different
phases of different solar cycles \citep{pevt08}.
We have studied forty three AR's (as shown in Table 1)
mostly observed during the declining phase of the solar
cycle 23.
None but five follow the twist hemispheric rule.

In the following section (Section 2), we discuss the
data sets used. Section 3 describes the analysis and results
obtained. Finally in Section 4 we present our conclusions.

\section{DATA SETS USED}
We have used the vector magnetograms obtained from Solar
Optical Telescope/Spectro-polarimeter
(SOT/SP: \cite {tsun08,suem08,ichi08})
onboard Hinode \citep{kosu07}
and the Advanced Stokes Polarimeter (ASP; \cite{elm92})
as well as the Diffraction Limited Spectro-polarimeter
(DLSP; \cite{sankar04,sankar06}) of the Dunn Solar
Telescope (DST). Standard and well-established
calibration procedure was adopted for ASP/DLSP data.
The procedure for obtaining the vector fields
from the ASP/DLSP data are described
elsewhere \citep{elm92,sankar04,sankar06}.

The Hinode (SOT/SP) data have been calibrated by the standard
``sp\_prep'' routine available in the Solar-Soft packages.
The prepared polarization spectra have been inverted to
obtain vector magnetic field components using an
Unno-Rachkowsky \citep{unno56,rach67}
inversion under the assumption of Milne-Eddington (ME)
atmosphere \citep{lando82,skum87}.
The 180$^o$ azimuthal ambiguity in the data sets are
removed by using acute angle method \citep{harv69,saku85,cupe92}.
All the data sets used, have high spatial sampling.
For example ASP $\sim0.3$ arcsec/pix, DLSP $\sim0.1$
arcsec/pixel and Hinode (SOT/SP) $\sim0.3$ arcsec/pixel.
However, a few observations are seeing limited to about an arcsec.

To minimize noise, pixels having transverse $(B_t)$ and
longitudinal magnetic field $(B_z)$ greater than a certain
level are only analyzed.
A quiet Sun region is selected for each sunspot and
1$\sigma$ deviation in the three vector field
components $B_x$, $B_y$ and $B_z$ are evaluated
separately. The resultant deviations in $B_x$ and $B_y$
is then taken as the 1$\sigma$ noise level
for transverse field components. Only
those pixels where longitudinal and transverse
fields are simultaneously greater than twice the
above mentioned noise levels are analyzed.

The data sets with their observation details are given
in Table 1.
The data sets observed from August 2001 to
April 2005 are obtained with the ASP and those
observed from June 2005 to December 2005
are from the DLSP.
Two vector magnetograms observed on 09 January 2007
and 06 February 2007 from Solar Vector
Magnetograph at Udaipur Solar Observatory
(SVM-USO: \cite{gosain04,gosain06}) and reported
in \cite{tiw08}, have also been included to improve
the statistics.
All the other data sets obtained since November 2006
onwards are taken from Hinode (SOT/SP).

\section{\textbf{DATA ANALYSIS AND RESULTS}}
We have used the following formula to compute the local
$\alpha$ values
\begin{equation}\label{}
  \alpha = \frac{(\nabla \times\bf B)_z} {B_z}     .
\end{equation}
The global $\alpha$ value of the active regions is estimated
from the following formula as described in \cite{tiw09}
\begin{equation}\label{}
\alpha_{g}=\frac{\sum(\frac{\partial B_y}{\partial x} -
\frac{\partial B_x}{\partial y})B_z}{\sum B_z^2}     .
\end{equation}

This estimate was shown to be not seriously affected
by the polarimetric noise \citep{tiw09}. Moreover,
since $\alpha_g$ is weighted by strong field values
\citep{hagi04} and not affected by singularities at
polarity inversion lines \citep{tiw09},
this parameter is more accurate
than a simple average of local $\alpha$.

\cite{hagy84} introduced the shear angle
$ \Delta\Phi = \Phi_{obs} - \Phi_{pot}$,
where $\Phi_{obs}$ and $\Phi_{pot}$ are the azimuthal
angles of observed and potential fields respectively.
The amplitude of this angle was studied
at the polarity inversion lines to investigate the flare
related changes \citep{hagy90,amba93,hagy99b}.
To emphasize the sign of shear angle
we wish to introduce the signed shear angle (SSA) for
the sunspots as follows :
We choose an initial reference azimuth for a
current-free field (obtained from
observed line of sight field).
Then we move to the observed field azimuth
from the reference azimuth through an acute angle.
If this rotation is counter-clockwise, then
we assign a positive sign for the SSA.
A negative sign is given for clockwise rotation.
This sign convention will be consistent with
the sense of azimuthal field
produced by a vertical current. This sign convention
is also consistent with the sense of chirality
(for details, see Appendix A).
The potential field has been computed using the
method of \cite{saku89}. Mean of the SSA obtained for
a whole sunspot is taken as the global value of
SSA for that sunspot.

The force-free parameter $\alpha$ involves three dimensions
since it basically represents the rate of change of rotation
per unit axial length.
The SSA is the rotational deviation of the
projection of the field onto the photosphere from that
of a reference current-free field.
The $\alpha$ parameter is a gradient of angle per unit
length while SSA is just an angle.
We therefore cannot expect a strong correlation between
the amplitudes of both the quantities, the SSA and the
$\alpha$ parameter.
But we do find a good correlation between their signs
as evident from Table 1.

Signed shear angle (SSA) provides the sign of twist
irrespective of whether the photospheric magnetic field
is force-free or not. Table 1 shows that the sign of
$\alpha$ is same as the sign of SSA.
Thus, we conclude that even if the photosphere is non
force-free, the sign of global $\alpha$ will empirically
give the sign of global SSA and therefore the sign of
global twist (chirality) of the sunspots.

To avoid any kind of projection effect we have
transformed the data sets to disk center \citep{venk89}
if the observed sunspot is equal to more than 10 degrees
away from the disk center.
In some active regions both the polarities
are compact enough to be studied separately.
We have treated each pole of those active regions
as an individual sunspot and this is denoted in the
Table 1 after the NOAA no. of sunspots by plus or
minus sign.

Two examples of the local $\alpha$ distribution for the
data sets obtained from Hinode (SOT/SP) are shown
in Figure 1.
The positive/negative contours are shown in red/blue
colors. The local $\alpha$ patches are seen in the umbra
and filamentary distribution of $\alpha$ is observed in
the penumbral region.
We find that the inclination angles oscillate between
$\sim$ 30-80 degrees when we go along azimuthal direction
in the filamentary penumbral structures.
This is consistent with the interlocking-comb
penumbral structure \citep{ichi07}
of the penumbral magnetic fields.
The vertical current J$_z$ has two components, viz.
$-\frac{1}{r}\frac{\partial B_r}{\partial\phi}$ ~and~
$\frac{1}{r}\frac{\partial (r B_{\phi})}{\partial r}$.
If we approximate the observed transverse field (B$_t$)
to be mostly radial (B$_t \sim B_r$) then we can
interpret the azimuthal variation of J$_z$ to result
from the term $-\frac{1}{r}\frac{\partial B_r}{\partial\phi}$.
This term is not expected to contribute to global
twist.
However $\frac{1}{r}\frac{\partial (r B_{\phi})}{\partial r}$
could be an important contributor to the global twist.
A detailed investigation of this interesting
possibility is deferred to another paper. For the
present, we obtain positive and negative values of
current side by side in the penumbra.
Because the $\alpha$ parameter depends on the current,
this oscillation in the filamentary structure
across the penumbral filaments is
expected for the $\alpha$ values too.

The distribution of vertical current and local $\alpha$
in the penumbra show higher
values than that in the umbral regions. An arc and a
straight line, selected respectively in the penumbra and umbra
of AR NOAA 10933 have been over plotted as shown in the
left panel of Figure 1. The corresponding values of vertical
current and $\alpha$ along the arc and the line are shown
in the Figure 2(a) and Figure 2(b) respectively.
We can see that both the positive and negative vertical
current as well as $\alpha$ are equally distributed in
the penumbra along the azimuthal direction.
This gives a negligible contribution to the global current
and global $\alpha$ values thereby indicating that the
contribution of
$\frac{1}{r}\frac{\partial (r B_{\phi})}{\partial r}$
is indeed small.
We have selected an arc rather than
the complete circle because many times sunspots are not
circular and therefore selecting a proper penumbral
region is not possible by a full circle.
Similar arcs have been selected in the other
sunspots and all the time it is seen that both the
positive and the negative vertical current as well as
$\alpha$ are distributed equally in the penumbra
giving negligible contribution to their
global values.
While current and $\alpha$ variations are correlated for
positive B$_z$, they will be anti correlated for
negative B$_z$.

Figure 2(b) shows a typical profile of spatial variations
of current and $\alpha$ across the umbra (along the line)
in the AR NOAA 10933 shown in the left panel of Figure 1.
We see that the amplitude of variation of $\alpha$ in the
umbra is smaller than that in the penumbra by
approximately an order of the magnitude and is of the
same order as that of the global $\alpha$ value of the
whole sunspot. The current values are not as much smaller
than that in the penumbra but are less balanced out than
in the penumbra.

In the right panel of Figure 1 an arc like structure
(partial ring)
with bunch of red contours (positive $\alpha$) can be
observed. This is opposite to that of the dominant
negative global $\alpha$ of the sunspot.
Such partial rings with opposite sign of the global value
are observed in 10 of the sunspots from our sample.
In the rest of the sunspots mixed current and $\alpha$
are present in the umbra with one dominant sign and
no such specific structures are seen at the
umbral-penumbral boundaries.

A few sunspots in our data sets studied
are small and have no penumbra.
Some ASP data doesn't show fine structures in penumbra
due to lack of spatial resolution.
We have included these active regions in our study
to look at the hemispheric behaviour of the global twist.

Most of the data sets we studied are observed during the
declining minimum phase of the solar cycle 23.
All except five of the sunspots observed do not
follow the twist hemispheric rule.

\section{DISCUSSION AND CONCLUSIONS}
We have introduced the concept of SSA for sunspots
and further find that the SSA and the global $\alpha$ value
of sunspots have the same sign.
Thus, $\alpha$ gives the same sign as the SSA and therefore
the same sign of the photospheric chirality of the
sunspots, irrespective of the force-freeness of
the sunspot fields.
As can be observed from the Table 1, the magnitudes
of SSA and $\alpha_{g}$ are not well correlated.
This lack of correlation could be due to a variety of
reasons: (a) departure from the force-free nature
(b) even for the force-free fields, $\alpha$ is
the gradient of twist variation whereas SSA is purely
an angle. The missing link is the scale length of
variation of twist. The magnitude of global SSA, therefore,
holds promise for characterizing global twist of the
sunspot magnetic fields, irrespective of its force-free
nature.

Patches of vertical current and $\alpha$ with opposite
signs are present in the umbra of each sunspot studied.
Since opposite currents repel, the existence of a
dominant current may be an useful binding force
for the umbra (cf. \cite{parker79}).
This will be examined in detail in further studies
of evolution of twist in decaying sunspots.
One sign of $\alpha$ dominates in the umbra which is also
seen to be the sign of the global $\alpha$ of the sunspot.
Similarly, the magnitude of the global $\alpha$ is
of the same order as the amplitude of the local $\alpha$
in the umbra.

The filamentary structures of vertical current and
local $\alpha$ are observed in the penumbra of the
sunspots and are, as discussed above, due to oscillatory
behaviour in the inclination values and therefore
gradients of the transverse field in the azimuthal
direction.
We find that the contributions of both
positive and negative values
of vertical current and $\alpha$ to their global
values cancel each other in the penumbra of the sunspot.
Thus the penumbral fine structures provide a negligible
contribution to the global $\alpha$ and current values of
sunspots.
The mutual repulsion of opposite currents also seem
to balance out in the azimuthal direction. It is
to be seen whether disruption of this balance leads
to sunspot rotation and change in global twist.
At any rate, the observed balancing
of the filamentary currents in the azimuthal direction
may be an important contribution to the force-free
nature of the sunspot fields.
The amplitude of $\alpha$ variation is approximately
an order of magnitude smaller in the umbral regions
than that in the penumbral regions and is of the
order of the global $\alpha$ of the whole sunspot.

Partial rings with opposite signs to that of
dominant sign of umbral $\alpha$ are observed
at sunspot umbral-penumbral boundaries in 10 out of 43
sunspots studied. However even in these few cases,
the rings are never complete in any sunspot.

Most of the AR's observed do not follow the twist
hemispheric rule. This issue of hemispheric rule needs
to be reinvestigated over a longer period as well as
with improved data.

Some researchers have tried to predict flare activity
from local distribution of $\alpha$
as well as global $\alpha$ values of sunspots
(\cite{nand08,hahn05} and references therein).
\cite{nand08} concluded from a study of AR 6982
that the global twist present
in a sunspot does not influence the flaring activity.
It is rather, governed by the spatial
distribution and evolution of twisted sub-structures
present in the sunspot. This conclusion indeed
needs more study.
We plan to address, in our forthcoming study,
the question of relation between flaring activity and
the role of global as well as the local
twist present in a large number of sunspots.

For the present, we demonstrate that the sign of
the SSA provides the sign of the photospheric
chirality irrespective of its force-free nature.
The sign of the global $\alpha$ of a sunspot is
determined by the dominant sign of umbral $\alpha$
values without much contribution from the
penumbral $\alpha$ values.

\acknowledgments {\bf Acknowledgements}\\
We thank the referee for very useful comments and suggestions.
Hinode is a Japanese mission developed and launched by ISAS/JAXA,
collaborating with NAOJ as a domestic partner, NASA and STFC (UK)
as international partners. Scientific operation of the Hinode
mission is conducted by the Hinode science team organized at
ISAS/JAXA. This team mainly consists of scientists from
institutes in the partner countries. Support for the post-launch
operation is provided by JAXA and NAOJ (Japan), STFC (U.K.),
NASA (U.S.A.), ESA, and NSC (Norway).
National Solar Observatory is Operated by the
Association of Universities for Research in
Astronomy (AURA) Inc., under cooperative
agreement with the National Science Foundation.

\appendix{APPENDIX}

\section{Relation between the sign of SSA and the sense of chirality}
The definition of the signed shear angle (SSA) is
introduced in the Section 3.
Figure 3 shows four structures, first two having
positive chirality and the next two having negative
chirality. The sign of B$_{pot}$ and B$_{obs}$ point
inward for negative B$_z$ and outward for positive B$_z$.
The rotation from B$_{pot}$ to B$_{obs}$ through
an acute angle is counter clockwise for the cases
of positive chirality and clockwise for negative
chirality. This is consistent with positive and
negative SSA respectively, by definition.
Thus, the sign of SSA will bear the
same sign of the chirality.


\begin{table}
\caption{List of the active regions studied. The global $\alpha$
value, the signed shear angle (SSA) and other details of the sunspots are
given:\label{tbl-1}}
\small{}
\centering
\begin{footnotesize}
\begin{tabular}{c c c c c c c}
\hline     
AR No.      &  Date of             &  Global Alpha         & Shear Angle               & Position      & Hemispheric\\
(NOAA)      &  Observation         &($\alpha_g$:/meter)    &(SSA: deg)        &               & Helicity Rule \\
\hline

10972      & 07 Oct 2007       & $-2.331\times 10^{-8}$      & $-1.085$                & S05W20(t)     & No  \\
10971	   & 29 Sep 2007	   & $3.053\times 10^{-8}$	     & $3.214 $	               & N03W07		   & No  \\
10970      & 05 Sep 2007       & $-2.001\times 10^{-8}$      & $-0.308$                & S07W58(t)     & No  \\
10969	   & 29 Aug 2007	   & $-3.424\times 10^{-8}$		 & $-4.488$	               & S05W33(t)	   & No  \\
10966	   & 09 Aug 2007	   & $-2.539\times 10^{-8}$	     & $-3.595$	               & S06E07	       & No  \\
10963($-$) & 12 Jul 2007	   & $-2.459\times 10^{-8}$	     & $-4.636$ 	           & S06E14(t)     & No  \\
10963($+$) & 12 Jul 2007       & $-3.440\times 10^{-8}$	     & $-4.495$       	       & S06E14(t)	   & No	 \\
10961	   & 02 Jul 2007       & $-5.119\times 10^{-8}$	     & $-4.973$	               & S10W16(t)	   & No	 \\
10960	   & 07 Jun 2007       & $3.027\times 10^{-8}$	     & $4.486$	               & S07W03	       & Yes \\
10956($-$) & 18 May 2007       & $9.642\times 10^{-8}$    	 & $11.595$	               & N02E07        & No  \\
10956($+$) & 18 May 2007       & $6.458\times 10^{-8}$		 & $5.352$	               & N02E07 	   & No	 \\
10955	   & 13 May 2007       & $-6.737\times 10^{-8}$	     & $-1.887$	               & S09W35(t)	   & No	 \\
10953	   & 29 Apr 2007	   & $-6.673\times 10^{-9}$	     & $-3.071$	               & S10E22(t)	   & No	 \\
10944      & 03 Mar 2007	   & $-2.084\times 10^{-8}$	     & $-4.635$	               & S05W30(t)	   & No	 \\
10941	   & 06 Feb 2007	   & $ -2.745\times 10^{-8}$	 & $-3.069$	               & S07W36(t)	   & No	 \\
10940	   & 01 Feb 2007	   & $-1.948\times 10^{-8}$	     & $-4.726$	               & S04W05        & No	 \\
10939($-$) & 23 Jan 2007	   & $-3.033\times 10^{-8}$	     & $-5.105$	               & S04W57(t)	   & No	 \\
10939(+)   & 23 Jan 2007 	   & $ -8.289\times 10^{-9}$	 & $-0.869$	               & S04W57(t)	   & No	 \\
10935 	   & 09 Jan 2007	   & $-2.412\times 10^{-8}$	     & $-3.414$	               & S07W30(t) 	   & No	 \\
10933	   & 05 Jan 2007	   & $ -1.119\times 10^{-9}$	 & $-2.423$	               & S04W01	       & No	 \\
10930($-$) & 12 Dec 2006       & $-3.519\times 10^{-8}$	     & $-6.676$	               & S05W21(t)	   & No	 \\
10930(+)   & 12 Dec 2006	   & $-1.624\times 10^{-7}$	     & $-18.067$	           & S05W21(t)	   & No	 \\
10926      & 03 Dec 2006       & $-7.049\times 10^{-9}$      & $-1.538$                & S09W32(t)     & No  \\
10923	   & 16 Nov 2006	   & $1.090\times 10^{-9}$	     & $0.350$	               & S05W30(t)	   & Yes \\
10921	   & 06 Nov 2006	   & $-3.318\times 10^{-7}$	     & $-14.054$	           & S08W38(t)	   & No	 \\
10841	   & 28 Dec 2005       & $1.114\times 10^{-7}$	     & $9.383$	               & N12E20(t)	   & No	 \\
10838	   & 22 Dec 2005       & $2.294\times 10^{-7}$	     & $14.757$	               & N17E20(t) 	   & No	 \\
10808($-$) & 13 Sep 2005	   & $1.017\times 10^{-7}$	     & $8.015$	               & S11E17(t)	   & Yes \\
10808(+)   & 13 Sep 2005	   & $1.225\times 10^{-7}$	     & $1.020$	               & S11E17(t)	   & Yes \\
10804	   & 26 Aug 2005	   & $-4.977\times 10^{-8}$	     & $-5.237$	               & N11W02	       & Yes \\
10803	   & 26 Aug 2005       & $2.559\times 10^{-7}$	     & $6.151$	               & N12E53(t)	   & No  \\
10800	   & 26 Aug 2005	   & $1.331\times 10^{-7}$	     & $3.967$	               & N17W49(t)	   & No  \\
10782	   & 02 Jul 2005	   & $-3.626\times 10^{-7}$	     & $-10.230$	           & S17W18(t)	   & No	 \\
10781	   & 04 Jul 2005	   & $1.027\times 10^{-7}$	     & $7.786$	               & N13W03 	   & No  \\
10780	   & 24 Jun 2005	   & $-6.357\times 10^{-8}$	     & $-0.806$	               & S08W28(t) 	   & No	 \\
10752	   & 17 Apr 2005       & $9.960\times 10^{-8}$       & $8.365$	               & N02W00 	   & No	 \\
10330	   & 09 Apr 2003	   & $3.988\times 10^{-8}$	     & $11.031$	               & N07W04 	   & No  \\
09601 	   & 03 Sep 2001	   & $1.367\times 10^{-8}$	     & $2.178$	               & N14W06(t) 	   & No	 \\
09596	   & 30 Aug 2001	   & $2.125\times 10^{-7}$	     & $9.297$	               & N21E15(t)	   & No	 \\
09591($-$) & 30 Aug 2001	   & $-2.359\times 10^{-7}$	     & $-6.111$	               & S18W36(t)	   & No	 \\
09591(+)   & 30 Aug 2001 	   & $-1.839\times 10^{-7}$	     & $-2.226$	               & S18W36(t) 	   & No  \\
09590	   & 26 Aug 2001	   & $-3.148\times 10^{-7}$	     & $-2.069$	               & S29W01(t) 	   & No	 \\
09585	   & 24 Aug 2001	   & $5.310\times 10^{-8}$	     & $1.730$	               & N14W30(t) 	   & No  \\

\hline
(t) :  \it transformed \\
\end{tabular}
\end{footnotesize}
\end{table}

\begin{figure}
\plottwo{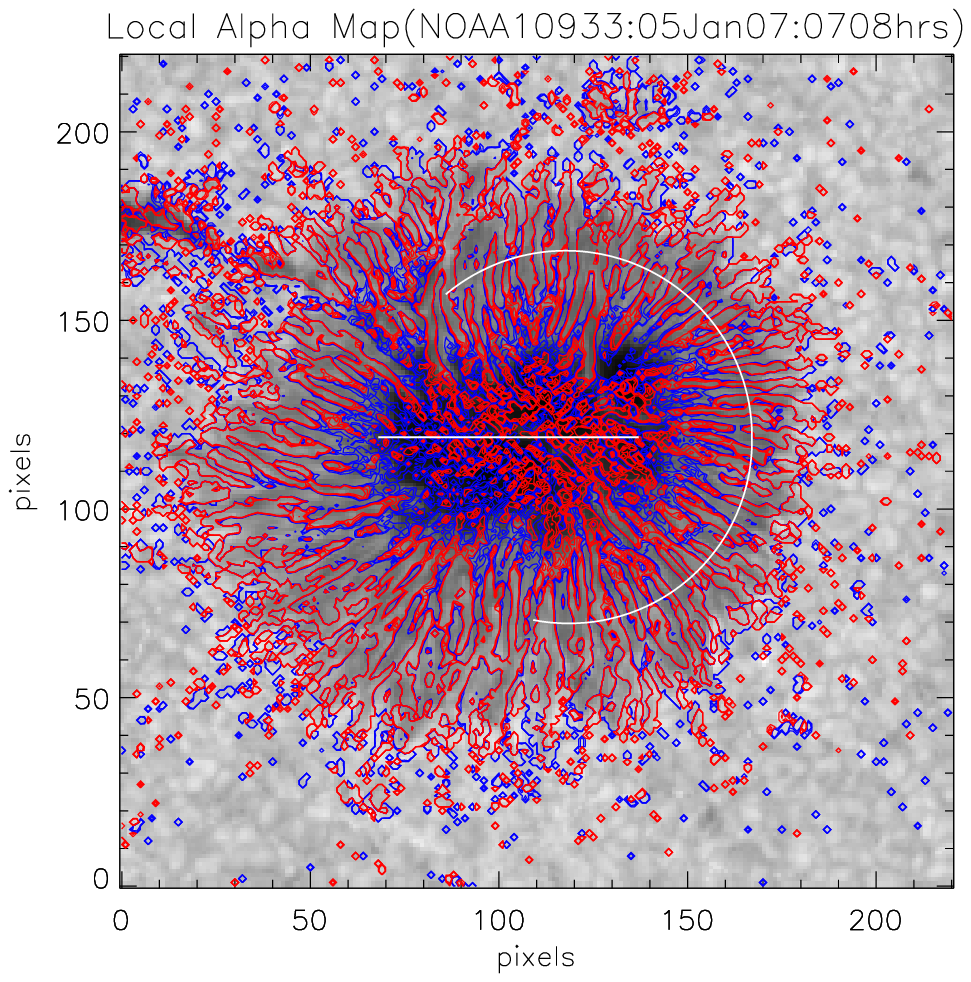}{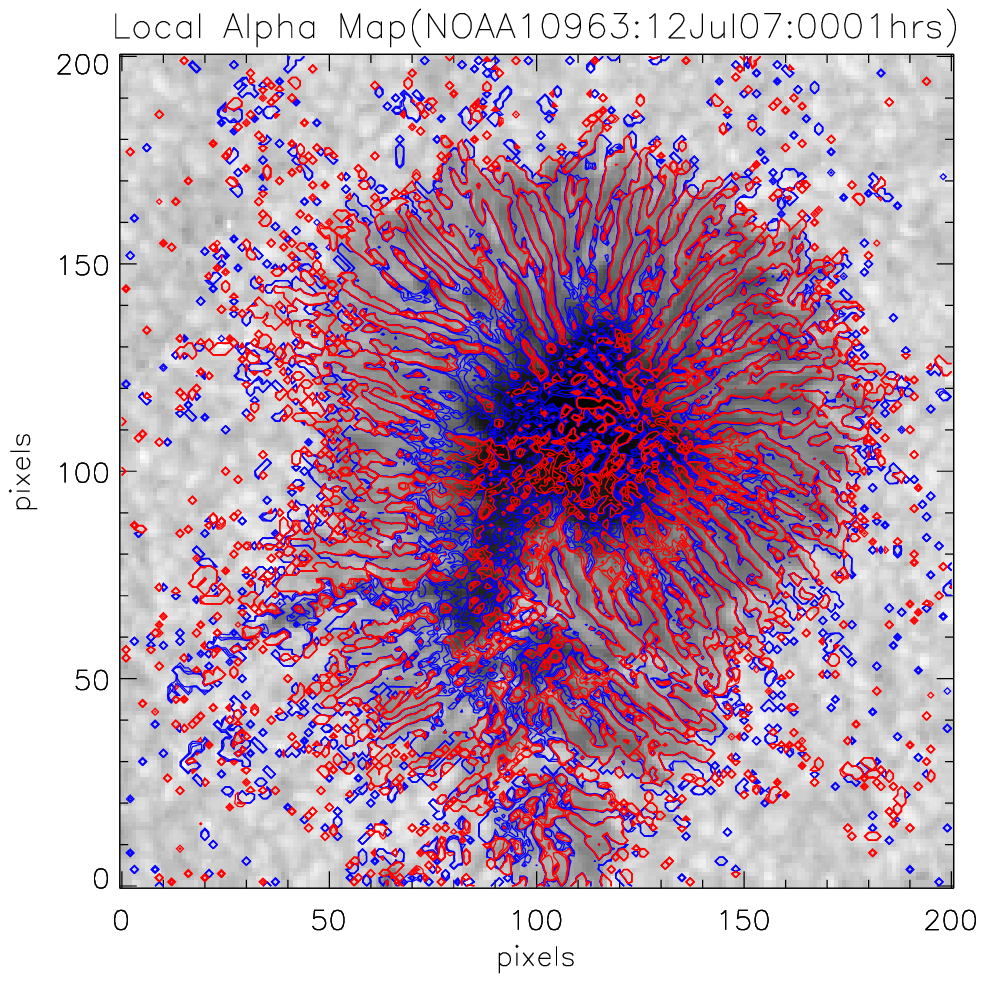}
\caption{Two examples of local $\alpha$ distribution
observed in Hinode (SOT/SP) data. Background is the continuum image.
Red and blue contours represent positive and negative values of
$\alpha$ respectively. The contour levels are $\pm 1\times 10^{-8} m^{-1},
\pm 5\times 10^{-8} m^{-1}$, and $\pm 10\times 10^{-8} m^{-1}$ .
The values of vertical current and $\alpha$ along the arc shown in the
penumbra of the image in the left panel, is plotted in Figure 2(a)
and those along the straight line in the umbra are plotted in
Figure 2(b).}
\end{figure}

\begin{figure}
\epsscale{.80}
\plotone{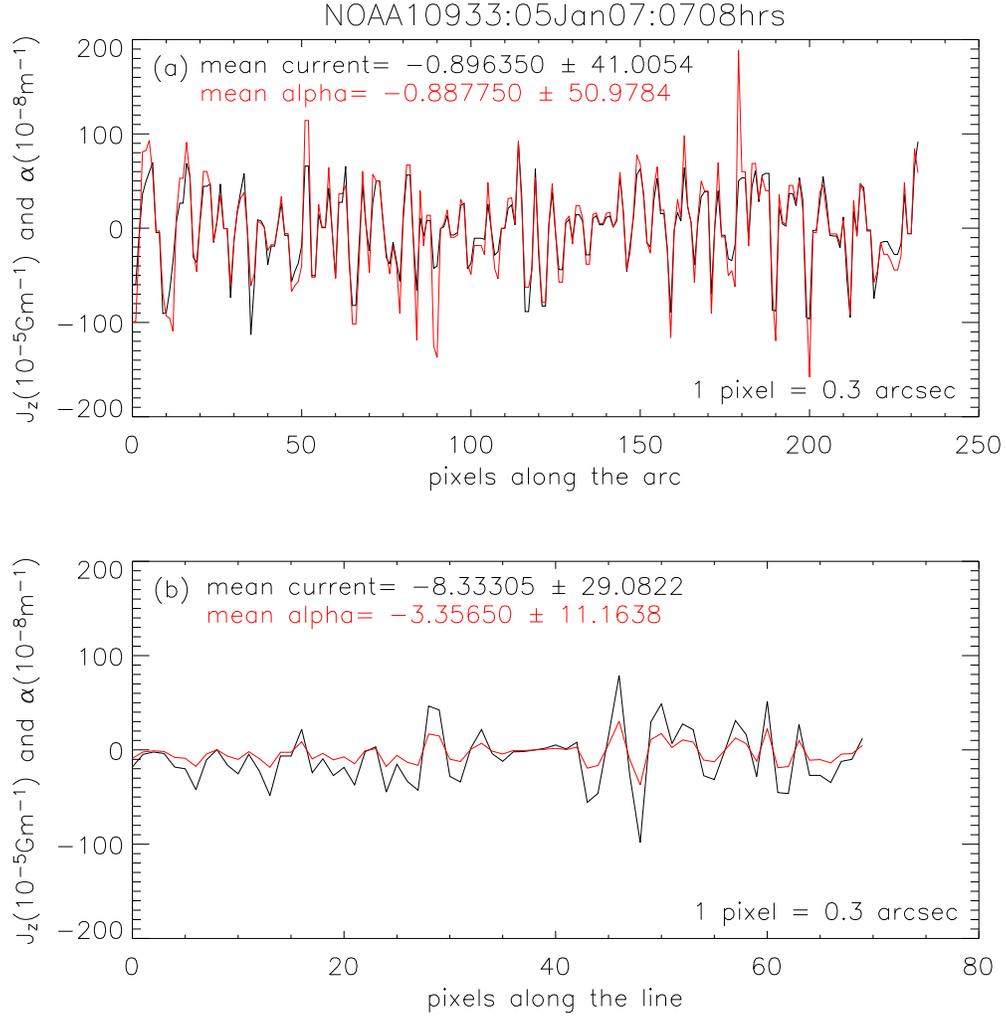}
\caption{Plots of vertical current and $\alpha$ values along
(a) the arc and (b) the straight line shown in the penumbra and
umbra of AR NOAA 10933 (left panel of Figure 1) respectively.
Black and red colors represent the current and
$\alpha$ values respectively.
The mean values of both the vertical current and
the $\alpha$ values with their 1$\sigma$ standard deviations
in the arc and the line are printed
on the plots in their respective colors.}
\end{figure}

\begin{figure}
\epsscale{.95}
\plotone{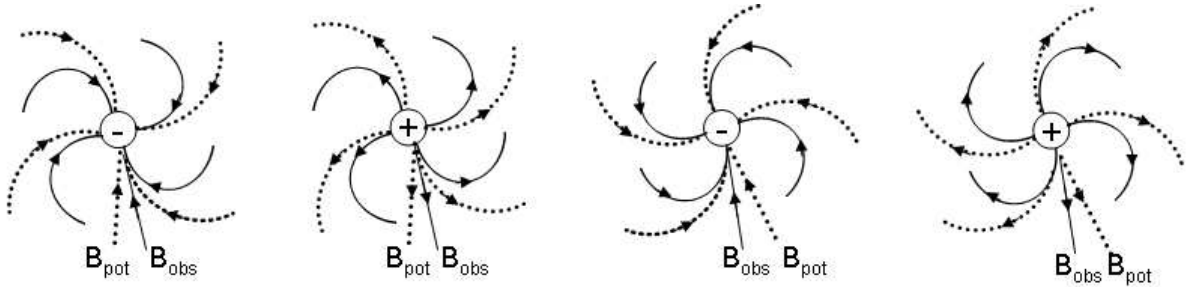}
\caption{Cartoon diagrams of circular unipolar spots with
positive and negative chirality are shown with
the directions of observed transverse field (B$_{obs}$) and
potential transverse field (B$_{pot}$). Solid and dashed
lines represent observed and potential fields respectively.
In this cartoon, the B$_{obs}$ is tangential to the solid
curved lines, while B$_{pot}$ is tangential to the dashed
curved lines which have lesser curvature than the solid
lines. First two cases bear positive chirality and later
two negative chirality. Plus and minus sign in the central
circular region represents the positive and negative
polarity respectively. For details see the text.}
\end{figure}

\end{document}